\documentclass[aps,prl,twocolumn,groupedaddress]{revtex4-1}

\usepackage{amsmath}
\usepackage{graphicx}
\usepackage{braket}

\begin{document}


\title{A quantum photonics model for non-classical light generation using integrated nanoplasmonic cavity-emitter systems}


\author{Fr\'ed\'eric Peyskens}
\email[]{fpeysken@mit.edu}
\affiliation{Quantum Photonics Group, RLE, Massachusetts Institute of Technology, Cambridge, Massachusetts 02139, USA}
\author{Dirk Englund}
\affiliation{Quantum Photonics Group, RLE, Massachusetts Institute of Technology, Cambridge, Massachusetts 02139, USA}


\date{\today}

\begin{abstract}
The implementation of non-classical light sources is becoming increasingly important for various quantum applications. A particularly interesting approach is to integrate such functionalities on a single chip as this could pave the way towards fully scalable quantum photonic devices. Several approaches using dielectric systems have been investigated in the past. However, it is still not understood how on-chip nanoplasmonic antennas, interacting with a single quantum emitter, affect the quantum statistics of photons reflected or transmitted in the guided mode of a waveguide. Here we investigate a quantum photonic platform consisting of an evanescently coupled nanoplasmonic cavity-emitter system and discuss the requirements for non-classical light generation. We develop an analytical model that incorporates quenching due to the nanoplasmonic cavity to predict the quantum statistics of the transmitted and reflected guided waveguide light under weak coherent pumping. The analytical predictions match numerical simulations based on a master equation approach. It is moreover shown that for resonant excitation the degree of anti-bunching in transmission is maximized for an optimal cavity modal volume $V_{c}$ and cavity-emitter distance $s$. In reflection, perfectly anti-bunched light can only be obtained for specific $(V_{c},s)$ combinations. Finally, our model also applies to dielectric cavities and as such can guide future efforts in the design and development of on-chip non-classical light sources using dielectric and nanoplasmonic cavity-emitter systems.
\end{abstract}


\maketitle


\section{Introduction}

It has been a long-standing goal in optical science to implement nonlinear effects at the few-photon level. In this regime, individual photons interact so strongly with one another that the propagation of light pulses containing few photons varies substantially with photon number. The strong dependence of light propagation on photon number for example allows the sorting or non-destructive counting of photons which could be used to implement various sources of non-classical light fields. \cite{ref8,ref1} One route to reach the quantum regime consists of coupling light to individual quantum emitters. \cite{refN1} The interaction between a single photon and a single quantum emitter is in general however relatively weak and hence makes the implementation of such quantum interconnects quite challenging. \cite{ref4} A particularly interesting approach is to route photons on photonic integrated circuits and interface them with on-chip quantum emitters as this allows integration of many functionalities on a single chip and hence paves the way towards truely scalable quantum devices. \cite{ref2,ref3} Several reports have investigated the interaction between quantum emitters and dielectric (Ref. \cite{refN4,refN5,refN3,refN6,ref5,refE3,refE2,refE1,ref9,ref10}) and plasmonic (Ref. \cite{refN2, ref6, ref11,ref12,ref23,ref14,ref24,ref25}) cavities and waveguides. Moreover, many theoretical investigations have been devoted to the interaction between quantum emitters and dielectric waveguides either without (Ref. \cite{refN4,refN5}) or with (Ref. \cite{ref4,refN3,refN6}) intermediate coupling to a dielectric cavity, as well as on emitters coupled to plasmonic cavities (Ref. \cite{ref23,ref14,ref24,ref25}) and waveguides (Ref. \cite{ref6,refN2,ref11}). Recently, the quantum statistics of photons scattered by a plasmonic nanocavity strongly coupled to a mesoscopic emitter ensemble was investigated under coherent pumping of the system. \cite{refOGV} However, the latter study did not incorporate plasmonic quenching nor the effect of coupling the cavity-emitter system to a nearby dielectric waveguide. As such, an outstanding question concerns how on-chip nanoplasmonic antennas, interacting with a single quantum emitter, affect the quantum statistics of photons reflected or transmitted in the guided mode of the waveguide. In this paper we will present a general quantum photonic model of evanescently coupled cavity-emitter systems and investigate the requirements for non-classical light generation using integrated nanoplasmonic cavities. Our model shows excellent correspondence with numerical simulations and as such allows accurate predictions of the photon statistics generated in the guided mode of a waveguide, coupled to a nanoplasmonic cavity-emitter system.

\section{Model}

\begin{figure}[ht]
\includegraphics[width=0.48\textwidth]{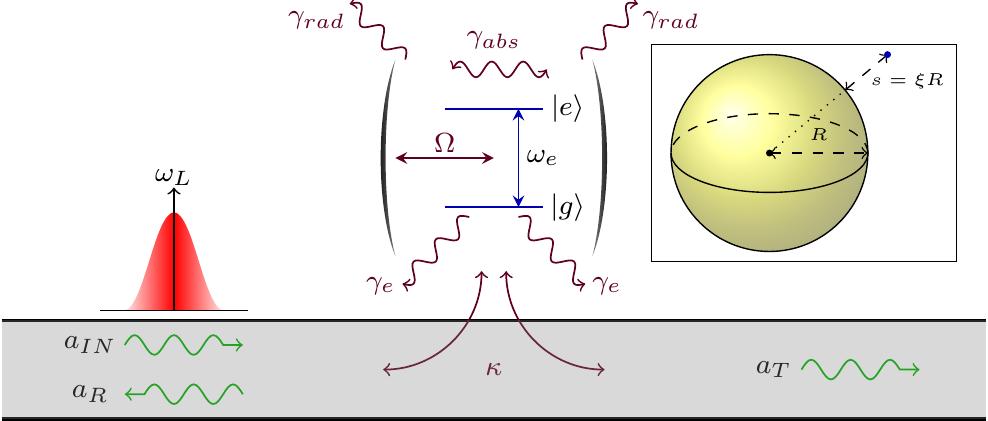}
\caption{\textbf{Quantum photonic platform.} The waveguide (gray) supports a 1D continuum of modes, carrying the excitation beam $a_{IN}$ (at frequency $\omega_{L}$) as well as the reflected ($a_{R}$) and transmitted ($a_{T}$) beam, and interacts evanescently with a cavity-emitter system. All coupling rates and frequencies are explained in the main text. The inset shows the case for a nanoplasmonic cavity-emitter system consisting of a spherical metallic nanoparticle with radius $R$ and an emitter at a distance $s=\xi R$ from the metal surface. \label{FigurePlatform}}
\end{figure}

The quantum photonic platform under investigation is shown in Fig. \ref{FigurePlatform}. It consists of a dielectric nanophotonic waveguide supporting a 1D continuum of left and right traveling modes which are characterized by the operators $l_{k}$ and $r_{k}$ respectively ($l_{k}$ and $r_{k}$ annihilate a left- or right- traveling photon with wavenumber $k=\omega_{k}/c$). These modes interact evanescently with a plasmonic cavity-emitter system, consisting of a cavity with resonance frequency $\omega_{c}$ (frequency of the fundamental mode), and a two-level quantum emitter which has a frequency difference $\omega_{e}$ between its ground $\ket{g}$ and excited $\ket{e}$ state ($S_{z}=1/2\left(\ket{e}\bra{e}-\ket{g}\bra{g}\right)$, $S_{+}=\ket{e}\bra{g}$, $S_{-}=\ket{g}\bra{e}$). The intrinsic cavity linewidth $\gamma_{c}$ is determined by the radiative decay rate to the non-guided modes ($\gamma_{rad}$) and the absorption decay rate ($\gamma_{abs}$), i.e. $\gamma_{c}=\omega_{c}/(2Q)=\gamma_{rad}+\gamma_{abs}$, where $Q$ is the unloaded quality factor. The overall linewidth $\gamma_{p}=\gamma_{c}+\kappa$ also incorporates the coupling rate $\kappa$ between the cavity mode (characterized by the operator $p$, annihilating a cavity excitation) and the waveguide mode. The coupling strength between the emitter and the cavity is given by $\Omega$. For dielectric cavities the emitter decay rate $\gamma_{e}$ is approximated by the decay rate of a dipole emitter in a background dielectric with index $n_{d}$, i.e. $\gamma_{e}=\gamma_{d}=\gamma_{0}\sqrt{\epsilon_{d}}=\gamma_{0}n_{d}$ with $\gamma_{0}=\frac{8\pi^{2}}{3\hbar\epsilon_{0}}\frac{|\textbf{d}|^{2}}{\lambda^{3}}$, where $\textbf{d}$ is the dipole moment vector of the emitter. For plasmonic cavities, $\gamma_{e}$ contains an additional contribution due to non-radiative quenching by higher order plasmon modes, as shown in Ref. \cite{refFP}, i.e. $\gamma_{e}=\gamma_{d}+\Omega^{2}f_{q}$ with
\begin{equation}
f_{q}=\sum_{l=2}^{\infty} \left(\frac{(2l+1)(l+1)^{2}}{12l (1+\xi)^{2l-2}}\right)\left(\frac{\gamma_{c}}{\left(\omega_{L}-\omega_{l}\right)^{2}+\left(\frac{\gamma_{c}}{2}\right)^{2}}\right),
\end{equation}
where $\omega_{l}$ is the resonance frequency of the $l-$th plasmon mode, assuming a spherical nanoplasmonic cavity with radius $R$ and an emitter positioned at a distance $s=\xi R$ from the metal surface, and $\omega_{L}$ the frequency by which the system is driven. For this particular case, the cavity-emitter coupling strength $\Omega=\sqrt{\frac{9 \pi c^{3}\gamma_{0}}{2\epsilon_{d}\omega_{p}^{2}(1+\xi)^{6}V_{c}}}$, while the coupling rate between the cavity and the waveguide is $\kappa=\frac{2\chi^{\kappa}\omega_{c}^{4}\epsilon_{d}\epsilon_{eff}V_{c}}{\pi^{2}c^{3}\epsilon_{wg}}$, with $\omega_{p}$ the plasma frequency of the metal, $\chi^{\kappa}$ a factor containing the overlap between the electric fields of the waveguide and the cavity, $\epsilon_{eff}$ the relative effective dielectric permittivity of the waveguide mode, $\epsilon_{wg}$ the relative permittivity of the waveguide core and $V_{c}=\frac{\pi R^{3}}{\epsilon_{d}}$ the cavity modal volume. \cite{refFP}
\\
The Hamiltonian of this quantum photonic platform is given by
\begin{multline}
\mathcal{H} =\hbar\omega_{e}S_{z} + \hbar\omega_{c}p^{\dagger}p + \hbar \Omega \left(pS_{+}+p^{\dagger}S_{-}\right)  \nonumber \\
+\hbar\int dk \omega_{k}l^{\dagger}_{k}l_{k} + \hbar \int dk \omega_{k}r^{\dagger}_{k}r_{k} +\mathcal{H}_{drive} \nonumber \\
+ \hbar g_{wg} \int dk\left(l^{\dagger}_{k}p + l_{k}p^{\dagger}\right) +\hbar g_{wg} \int dk\left(r^{\dagger}_{k}p + r_{k}p^{\dagger}\right).
\end{multline}
It includes the free Hamiltonian of the emitter, cavity and waveguide modes as well as the interaction between the emitter and the cavity and the interaction between the cavity and waveguide modes. The coupling constant between the cavity and the waveguide modes is $g_{wg}=\sqrt{\frac{c\kappa}{4\pi}}$. \cite{refN5} Finally, $\mathcal{H}_{drive}$ represents the coherent input driving field,
\begin{equation}
\mathcal{H}_{drive}=\hbar F p^{\dagger}\text{e}^{-i\omega_{L}t} + \hbar \bar{F}p \text{e}^{i\omega_{L}t}
\end{equation}
oscillating at a frequency $\omega_{L}$. The coherent driving field strength $F$ can be related to the input field $a_{IN}$ by considering that the coupling between the continuum of forward (right) propagating waveguide modes and the cavity is given by
\begin{equation}
\hbar g_{wg}\int dk r^{\dagger}_{k}p + \text{h.c.} \equiv \hbar \bar{F}p + \text{h.c.} \Rightarrow F=g_{wg}\int dk r_{k}(t_{0})
\end{equation}
and that the input field $a_{IN}$ is just the initial field at $t=t_{0}$, i.e. $a_{IN}=\sqrt{2\pi}^{-1}\int dk r_{k}(t_{0})$, which can be obtained by formally solving the Heisenberg equations for the system operators. Hence $a_{IN}=\frac{F}{\sqrt{2\pi}g_{wg}}=\sqrt{\frac{2}{c\kappa}}F$. To incorporate all loss channels, we model the overall platform using a master equation in a frame rotating at $\omega_{L}$  (see Supplementary Information)
\begin{align}
\frac{d\rho}{dt} &= -\frac{i}{\hbar}[\mathcal{H}_{rot},\rho] + \frac{\gamma_{p}}{2}\left(2 p\rho p^{\dagger} - p^{\dagger}p\rho - \rho p^{\dagger}p \right) + \nonumber \\ &\ \ \ \ \ \ \ \frac{\gamma_{e}}{2}\left(2 S_{-}\rho S_{+} - S_{+}S_{-}\rho - \rho S_{+}S_{-} \right)
\end{align}
where $\mathcal{H}_{rot}=\hbar\delta_{e}S_{z}+\hbar\delta_{c}p^{\dagger}p+\hbar \Omega \left(pS_{+}+p^{\dagger}S_{-}\right)+\hbar F p^{\dagger}+ \hbar \bar{F}p$ with $\delta_{e}=\omega_{e}-\omega_{L}$ and $\delta_{c}=\omega_{c}-\omega_{L}$.  
\par
When solving the stationary Heisenberg equations in a frame rotating at the drive frequency $\omega_{L}$ (Ref. \cite{refN4,refN3}), the average value of the operators ($\langle\Psi|p|\Psi\rangle\overset{\Delta}{=}p$, $\langle\Psi|S_{-}|\Psi\rangle\overset{\Delta}{=}s_{-}$, $\langle\Psi|S_{z}|\Psi\rangle\overset{\Delta}{=}s_{z}$) is approximately given by
\begin{align}
	p&=-i\Omega\tau s_{-} - i\tau\sqrt{\frac{c\kappa}{2}}a_{IN} \\
	s_{-}&= \frac{\beta\tau\Omega\sqrt{2c\kappa}a_{IN}}{1+\Omega^{2}\beta\tau}s_{z} \\
	s_{z}&=-\frac{1}{2}\left(\frac{1}{1+\frac{|a_{IN}|^{2}}{P_{c}}}\right)=-\frac{1}{2}\left(\frac{1}{1+x}\right)
\end{align}
with $\frac{1}{\tau}=i(\omega_{c}-\omega_{L})+\frac{\kappa}{2}+\frac{\gamma_{c}}{2}=i\delta_{c}+\frac{\gamma_{p}}{2}$, $\frac{1}{\beta}=i(\omega_{e}-\omega_{L})+\frac{\gamma_{e}}{2}=i\delta_{e}+\frac{\gamma_{e}}{2}$, $A=\frac{\beta\tau\Omega\sqrt{2c\kappa}}{1+\Omega^{2}\beta\tau}$ and
\begin{equation}
P_{c}=\frac{\gamma_{e}}{\Omega\sqrt{2c\kappa}\Im\left(i\tau \overline A\right)-\Omega^{2}\Re\left(\tau\right)\left|A\right|^{2}}.
\end{equation}
Since the cavity is evanescently coupled to the waveguide modes, the reflected $a_{R}$ and transmitted $a_{T}$ fields are given by
\begin{align}
	a_{R}&=-i\sqrt{\frac{\kappa}{2c}}p \\
	a_{T}&=a_{IN}-i\sqrt{\frac{\kappa}{2c}}p.
\end{align}
The power reflection and transmission coefficients are respectively given by $\mathcal{R}=|r|^{2}=|\frac{a_{R}}{a_{IN}}|^{2}$ and $\mathcal{T}=|\frac{a_{T}}{a_{IN}}|^{2}=|1+r|^{2}$. In the remainder of this paper we assume to work in the weak probe limit, i.e. when the input power is sufficiently small such that $|a_{IN}|^{2}\ll P_{c}$ implying $x\approx 0$. In this limit, we can derive simple analytical approximations for the second order correlation functions $g^{(2)}_{\mathcal{T/R}}$, as will be shown further on. For all numerical calculations, we assume a fixed coherent coupling strength $F=10^{-5}\omega_{c}$ in order to operate in the weak probe limit (in this paper $\omega_{c}\approx 2960$ THz corresponding to a free space wavelength of 637 nm).

\begin{figure*}[ht]
\includegraphics[width=\textwidth]{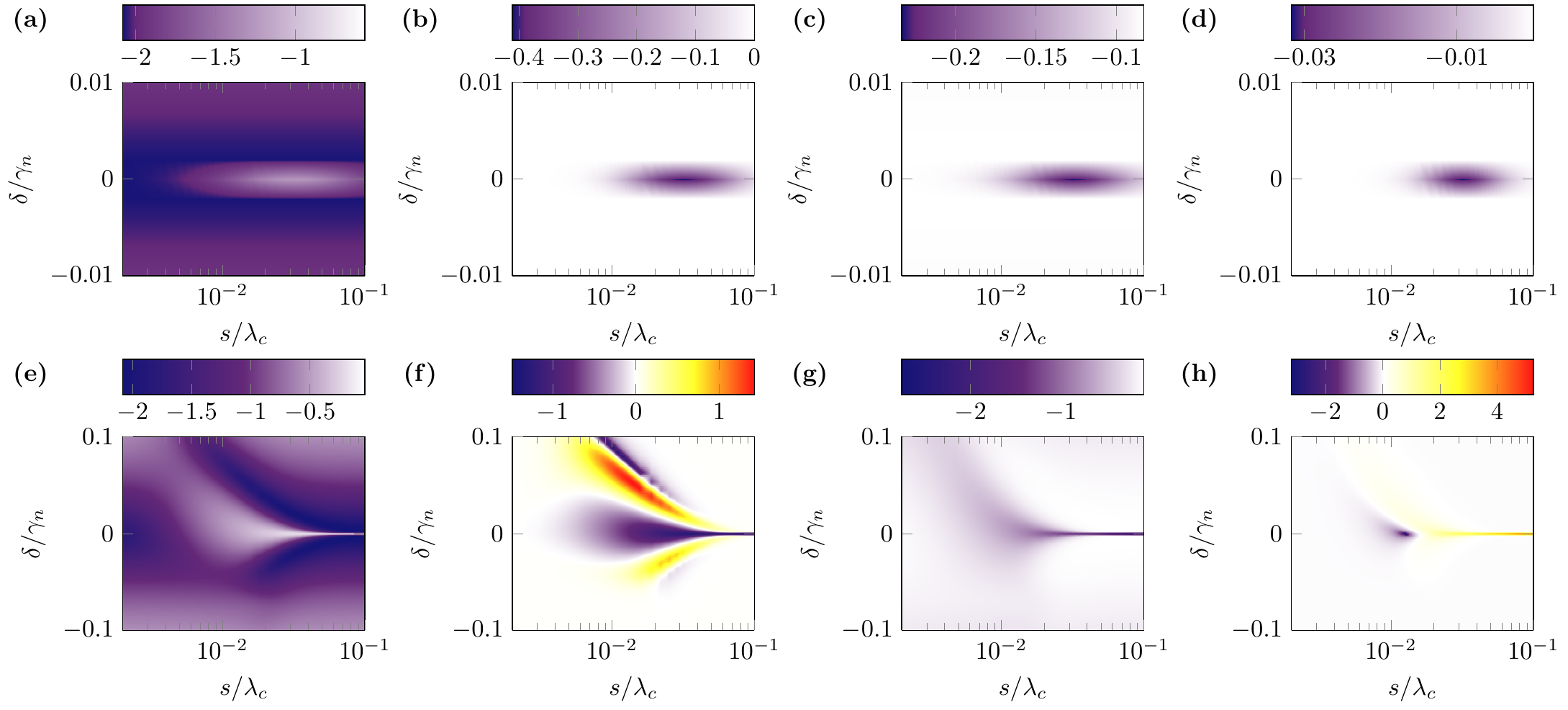}
\caption{
Properties of the transmitted \textbf{(a-b)/(e-f)} and reflected \textbf{(c-d)/(g-h)} guided light for \textbf{(a-d)} an integrated spherical nanoparticle with $Q=15$ at $\lambda_{c}=637$ nm, $\kappa=10\gamma_{c}$, $\gamma_{D}=1$ GHz, $\epsilon_{eff}=2.56$, $\epsilon_{wg}=4$ (common values for a SiN waveguide), $\chi^{\kappa}=1$ and \textbf{(e-h)} a nanoplasmonic cavity with an increased polarizability and field enhancement (as explained in the main text). \textbf{(a/e)} $\mathcal{T}$  and \textbf{(b/f)} $g^{(2)}_{\mathcal{T}}(0)$ as a function of $\delta_{c}=\delta_{e}=\delta$ ($\gamma_{n}=3\gamma_{p}$) and $s$. \textbf{(c/g)} $\mathcal{R}$  and \textbf{(d/h)} $g^{(2)}_{\mathcal{R}}(0)$ as a function of $\delta$ and $s$. In order to incorporate quenching, we used 1000 higher order modes. All plots are on a $\log_{10}$ scale.
\label{FigureNZD}}
\end{figure*}

\section{Discussion}

Using the above model, we now investigate the properties of the transmitted and reflected guided light for a waveguide evanescently coupled to a spherical metallic nanoparticle (Fig. \ref{FigureNZD}). We assume that the cavity is perfectly tuned with the emitter ($\delta_{c}=\delta_{e}=\delta$). The coupling rate between the cavity and the waveguide is fixed to $\kappa=10\gamma_{c}$ (guaranteeing a coupling efficiency of $\kappa/(\kappa+\gamma_{c})\approx91\%$). Since $\kappa$ is fixed, the cavity modal volume is fixed as well. Figures \ref{FigureNZD}(a)/(b) and Figures \ref{FigureNZD}(c)/(d) respectively show $\mathcal{T}$/$g^{(2)}_{\mathcal{T}}(0)$ and $\mathcal{R}$/$g^{(2)}_{\mathcal{R}}(0)$  as a function of the detuning $\delta$ and emitter-cavity distance $s$ (which basically tunes the cavity-emitter coupling strength as $V_{c}$ is already fixed by fixing $\kappa$). These plots show that both for the transmitted and reflected case, anti-bunched light is only observed near $\delta=0$. As opposed to dielectric cavities, where a strong anti-bunching in transmission appears at one of the normal mode splitted frequencies $\omega_{L}=\omega_{c}\pm\delta_{d}$ ($\delta_{d}\neq 0$), no photon blockade effect appears for this system. This result stems from the fact that for the given $\kappa$ (which needs to exceed $\gamma_{c}$), $\Omega$ can not be made large enough. Indeed, since $\kappa>\gamma_{c}$ to observe significant anti-bunching in transmission, the radius of the nanoparticle (and by extension modal volume) needs to be large enough, which in turn makes $\Omega$ smaller. One way to maintain a large $\kappa$ without enlarging the modal volume would be to increase the cavity polarizability, e.g. by switching from spherical nanoparticles to rod or bowtie antennas. The use of such cavities would also result in an increased field enhancement (with respect to a single spherical nanoparticle) in the gap near the metal surface. If one used a nanoplasmonic cavity with cavity polarizability 10 times larger than the one for a spherical particle, while keeping $\kappa=10\gamma_{c}$ (implying a decrease in the modal volume by 100), and for which field enhancement effects further increase $\Omega$ by a factor 100, this would result in an off-resonant anti-bunching in transmission as shown in Fig.\ref{FigureNZD}(e-f). However, the photon blockade effect is in this case asymmetric with respect to $\delta=0$ due to the Lorentzian lineshape functions appearing in $f_{q}$. Photon blockade will at first only appear when the probe wavelength is redshifted (i.e. $\delta>0$) compared to the resonance wavelength of the fundamental plasmonic mode. Since $\{\omega_{l}-\omega_{1}\geq 0,\forall l\}$, $\gamma_{e}$ will first increase if $\delta<0$ due to the reduced denominator in the Lorentzian lineshape function of $\gamma_{e}$. Therefore a certain minimum $\Omega$ is required to compensate for the increased quenching effect in the relevant $\delta<0$ region. It is also clear that the transmission of the strongly off-resonant anti-bunched light is much smaller than the transmission of the resonant anti-bunched light. Potential plasmonic cavities that allow increased polarizability and field enhancement are e.g. double rod or bowtie antennas. \cite{ref22,ref21} Moreover, quenching could be reduced by considering hybrid metal-dielectric cavities, as recently studied in Ref. \cite{refAK}. In reflection, a negligible anti-bunching is observed for the spherical nanoplasmonic cavity and the given $\kappa$. Since the anti-bunching in this case solely stems from the light emitted by the quantum emitter and not from an interference between the excitation beam and the emitted light, the conditions are different. In reflection, $\kappa$ may be smaller than $\gamma_{c}$ to obtain strong resonant anti-bunching, as long as $\Omega$ satisfies $\Omega=\sqrt{\gamma_{e}\left(\gamma_{e}+\gamma_{p}\right)}/2$ (see below). This condition results in a quadratic equation for $\Omega^{2}$ (because $\gamma_{e}$ explicitly depends on $\Omega^{2}$ through $\gamma_{e}=\gamma_{d}+\Omega^{2}f_{q}$), yielding two specific solutions for $\Omega$ around which strong resonant anti-bunching is obtained (even if $\kappa<\gamma_{c}$). 
\par
\begin{figure}[ht]
\includegraphics[width=0.48\textwidth]{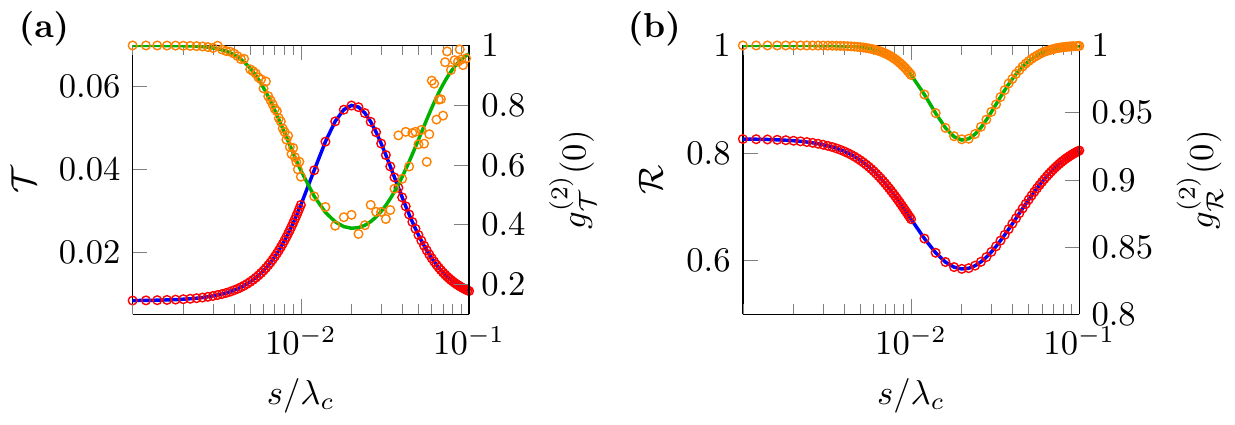}
\caption{\textbf{(a)} Numerically calculated (dots) and theoretically predicted (lines) values of $\mathcal{T}$ (left axis, red dots and blue line) and $g^{(2)}_{\mathcal{T}}$ (right axis, orange dots and green lines) for $\delta=0$. \textbf{(b)} Same as \textbf{(a)} but for the reflected field. The other parameter values are the same as for Fig. \ref{FigureNZD}. \label{FigureZDcomp}}
\end{figure}
We will now investigate the case of resonant excitation in more detail. Since we assume to work in the weak probe limit, the reflection and transmission coefficients are given by:
\begin{align}
\mathcal{R}&=\left|\frac{\kappa}{\gamma_{p}(1+\mathcal{C})}\right|^{2} \\
\mathcal{T}&=\left|\frac{\gamma_{c}+\mathcal{C}\gamma_{p}}{\gamma_{p}+\mathcal{C}\gamma_{p}}\right|^{2},
\end{align}
where $\mathcal{C}=\frac{4\Omega^{2}}{\gamma_{p}\gamma_{e}}$ is the cooperativity of the system. In this limit, the Hilbert space used to evaluate the master equation can be truncated to states which have a maximum of two excitations (either emitter or cavity excitations), allowing us to derive analytical formulas for the second order correlation function $g^{(2)}(0)$ as a measure for the degree of anti-bunching. In the Supplementary Information it is shown that in this limit the $g^{(2)}_{\mathcal{T/R}}(0)$ functions on resonance are respectively given by
\begin{equation}
g^{(2)}_{\mathcal{T}}(0)=\frac{\left(1+\mathcal{C}\right)^{2}\left(1+\frac{4\Omega^{2}\left(\gamma_{c}+2\left(\gamma_{e}+\kappa\right)\right)}{\gamma_{c}\gamma_{e}\left(\gamma_{e}+\gamma_{p}\right)}+\frac{16\Omega^{4}}{\gamma_{c}^{2}\gamma_{e}\left(\gamma_{e}+\gamma_{p}\right)}\right)^{2}}{\left(1+\frac{\gamma_{p}}{\gamma_{c}}\mathcal{C}\right)^{4}\left(1+\frac{\gamma_{e}}{\gamma_{e}+\gamma_{p}}\mathcal{C}\right)^{2}}
\end{equation}
and
\begin{equation}
g^{(2)}_{\mathcal{R}}(0)=\frac{\left(4\Omega^{2}+\gamma_{e}\gamma_{p}\right)^{2}\left(\gamma_{e}\left(\gamma_{e}+\gamma_{p}\right)-4\Omega^{2}\right)^{2}}{\gamma_{e}^{4}\left(4\Omega^{2}+\gamma_{p}\left(\gamma_{e}+\gamma_{p}\right)\right)^{2}}.
\end{equation}
Figures \ref{FigureZDcomp}(a) and (b) respectively show the properties of the transmitted and reflected light. The red and orange dots present the respective numerically calculated $\mathcal{T}/\mathcal{R}$ and $g^{(2)}_{\mathcal{T}}(0)/g^{(2)}_{\mathcal{R}}(0)$ while the blue and green solid lines give the respective analytical predictions of $\mathcal{T}/\mathcal{R}$ and $g^{(2)}_{\mathcal{T}}(0)/g^{(2)}_{\mathcal{R}}(0)$. Our data show a perfect correspondence between the analytically predicted and numerical values. As such the analytical formulas derived here can be used to assess the degree of anti-bunching, in the weak probe limit, for any cavity-emitter system evanescently coupled to a waveguide (the formulas for off-resonant pumping are given in the Supplementary Information as well). It should be noted that these formulas hold for any $Q$ cavity, both dielectric and plasmonic.
\begin{figure*}[ht]
\includegraphics[width=\textwidth]{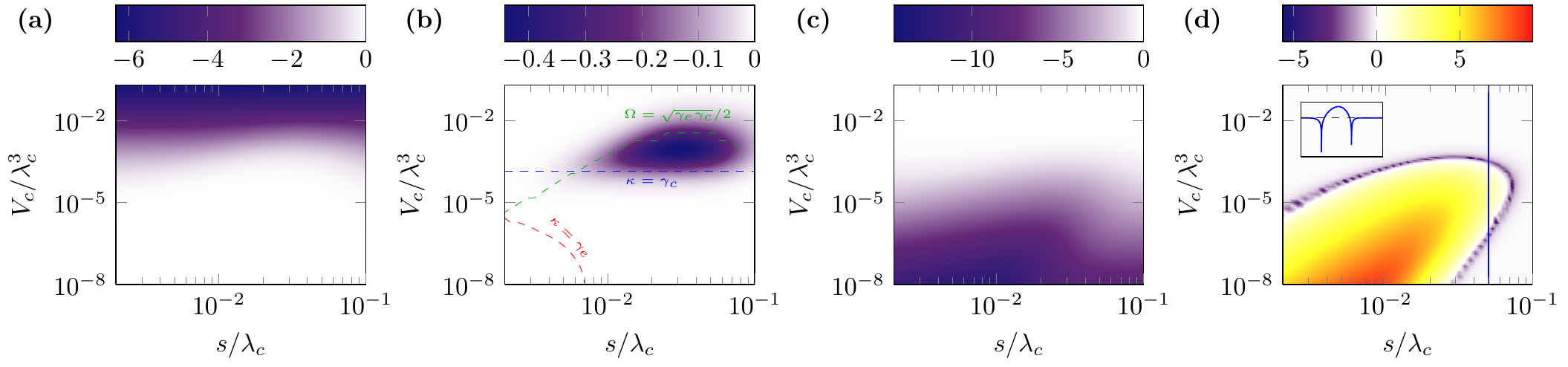}
\caption{Properties of the transmitted and reflected light for resonant excitation. \textbf{(a)} $\mathcal{T}$ and \textbf{(b)} $g^{(2)}_{\mathcal{T}}(0)$. The blue dashed line marks $\kappa=\gamma_{c}$, the red dashed line $\kappa=\gamma_{e}$ and the green dashed line $\Omega=\sqrt{\gamma_{e}\gamma_{c}}/2$. \textbf{(c)} $\mathcal{R}$ and \textbf{(d)} $g^{(2)}_{\mathcal{R}}(0)$. The inset in \textbf{(d)} is a slice of $g^{(2)}_{\mathcal{R}}(0)$ along a fixed $s$. All plots are on a $\log_{10}$ scale. \label{FigureZDD}}
\end{figure*}
\par
Figure \ref{FigureZDD} shows the properties of the transmitted and reflected light as a function of $V_{c}$ ($\kappa$) and $s$ ($\Omega$ for fixed $V_{c}$). It is clear that in transmission, a maximum degree of anti-bunching is obtained for an optimal cavity modal volume $V_{c}$ and cavity-emitter distance $s$. Moreover, anti-bunching can in general only be obtained when $\kappa>\text{max}(\gamma_{c},\gamma_{e})$. From the figure this is obvious for $\kappa>\gamma_{c}$. However it also holds for $\kappa>\gamma_{e}$, e.g. for dielectric cavities with very high $Q$ it is possible that $\gamma_{e}>\gamma_{0}$. For plasmonic cavities, the modal volume will generally be $Q-$factor limited due to the very high $\gamma_{c}$. As such there is a minimum modal volume $V_{min}$ for the transmitted light to be anti-bunched, i.e.
\begin{equation}
V_{c}>V_{min}=\lambda_{c}^{3}\left(\frac{\epsilon_{wg}}{32\pi Q\chi^{\kappa}\epsilon_{d}\epsilon_{eff}}\right).
\end{equation}
Interestingly, this lower bound $V_{min}$ equals to the optimum modal volume required for single photon generation when the system is initially pumped to the excited state and the subsequent decay of the emitted light is observed. \cite{refFP} On the other hand, $\Omega$ should also be large enough to observe anti-bunching, as evidenced by the green dashed line in Fig. \ref{FigureZDD}(b). When $\Omega$ becomes significantly smaller than $\sqrt{\gamma_{e}\gamma_{c}}/2$, the light becomes coherent again. Thus there is also an upper bound for the cavity modal volume $V_{max}$, depending on the quenching factor $f_{q}$, i.e.
\begin{equation}
V_{c}<V_{max}=\lambda_{c}^{3}\left(\frac{9Q(4-\gamma_{c}f_{q})}{8\pi^{2}\epsilon_{d}(1+\xi)^{6}}\right)\left(\frac{\gamma_{0}\omega_{c}^{2}}{\gamma_{d}\omega_{p}^{2}}\right).
\end{equation}
As a result, similar to the case of pure single photon generation (\cite{refFP}), it is only possible to generate non-classical light in transmission if the modal volume stays within the given bounds. As already highlighted in Ref. \cite{refFP}, it is hence not always beneficial to aim for a minimal cavity modal volume for non-classical light generation. Despite the fact that the degree of anti-bunching can be optimized, it is nevertheless impossible to achieve perfectly anti-bunched light in transmission. Bunching on resonance is also not observed in transmission for the investigated platform.
\par
However, in reflection, the situation is different and perfectly anti-bunched light can be achieved for $\Omega_{\mathcal{R}}^{opt}=\sqrt{\gamma_{e}\left(\gamma_{e}+\gamma_{p}\right)}/2$. The reflected light only exhibits non-classical behaviour in a very narrow region around $\Omega_{\mathcal{R}}^{opt}$, requiring a precise alignment of the quantum emitter with the cavity (Fig. \ref{FigureZDD}(d)). Within the boundary set by $\Omega_{\mathcal{R}}^{opt}=\sqrt{\gamma_{e}\left(\gamma_{e}+\gamma_{p}\right)}/2$, strong bunching is observed for the smaller modal volumes and cavity-emitter distances. Outside this boundary, the light is perfectly coherent. The overall reflection of non-classical light is however relatively low (Fig. \ref{FigureZDD}(c)). Evaluating $\mathcal{R}$ at $\Omega_{\mathcal{R}}^{opt}$ yields $\mathcal{R}_{opt}=\left(\frac{\kappa}{\gamma_{e}+2\gamma_{c}+2\kappa}\right)^{2}$. For $\kappa>\text{max}\left(\gamma_{e},\gamma_{c}\right)$ the maximum reflection of a non-classical light state is achieved and equals $25\%$.

\section{Conclusion}

We presented a general quantum photonic model of evanescently coupled cavity-emitter systems and investigated the requirements for non-classical light generation using integrated nanoplasmonic cavities. We considered a spherical metallic nanoparticle as a model system for the nanoplasmonic cavity because analytical formulas can be derived for this specific case. However, the model is generally applicable to any plasmonic (and dielectric) cavity, where the system parameters then need to be obtained from electromagnetic simulations. We derived analytical formulas for the first and second order correlation function in the weak probe limit, both for transmission and reflection, which showed excellent correspondence with numerical simulations using a quantum master equation approach. Moreover, we showed that non-classical light generation for the considered nanoplasmonic cavity-emitter system is only possible using resonant excitation and that photon blockade effects only appear for nanoplasmonic cavities with increased polarizability and field enhancement compared to a spherical metallic nanoparticle. This result stems from the intrinsic dependence of the effective emitter decay rate on the cavity-emitter coupling strength, due to quenching effects near the plasmonic nanoparticle. Investigation of resonant excitation in more detail showed that in transmission the degree of anti-bunching is maximized for an optimal cavity modal volume $V_{c}$ and cavity-emitter distance $s$. In reflection, perfectly anti-bunched light can be obtained for specific $(V_{c},s)$ combinations. The presented model allows accurate predictions of the photon statistics generated in the guided mode of a waveguide by an integrated nanoplasmonic cavity-emitter system, driven by a weak coherent probe beam. These results inform future efforts in the design and development of on-chip non-classical light sources.

\section*{Funding Information}
F.P. acknowledges support from a BAEF (Belgian American Educational Foundation) and Fulbright postdoctoral fellowship.

\section*{Acknowledgments}

We acknowledge Prof. Darrick Chang (ICFO) for fruitful discussions on the model and its implementation.

\section{Supplemental material}

\subsection{Master equation derivation and numerical evaluation}
\subsubsection{Derivation}
The waveguide supports a 1D continuum of left and right traveling modes which are characterized by the operators $l_{k}$ and $r_{k}$, which respectively annihilate a left- or right- traveling photon with wavenumber $k=\omega_{k}/c$. The Hamiltonian of the quantum photonic platform is then given by
\begin{multline}
\mathcal{H} =\hbar\omega_{e}S_{z} + \hbar\omega_{c}p^{\dagger}p + \hbar \Omega \left(pS_{+}+p^{\dagger}S_{-}\right)  \\
+\hbar\int dk \omega_{k}l^{\dagger}_{k}l_{k} + \hbar \int dk \omega_{k}r^{\dagger}_{k}r_{k} + \mathcal{H}_{drive}  \\
+ \hbar g_{wg} \int dk\left(l^{\dagger}_{k}p + l_{k}p^{\dagger}\right) +\hbar g_{wg} \int dk\left(r^{\dagger}_{k}p + r_{k}p^{\dagger}\right).
\end{multline}
It includes the free Hamiltonian of both the emitter, the cavity and the waveguide modes as well as the interaction between the emitter and the cavity and the interaction between the cavity and waveguide modes. The coupling constant between the cavity and the waveguide modes is $g_{wg}=\sqrt{\frac{c\kappa}{4\pi}}$. \cite{ref20} The term $\mathcal{H}_{drive}=\hbar F p^{\dagger}+ \hbar \bar{F}p$ represents the coherent driving field with strength $F$. The Heisenberg equation for the $r_{k}$ and $l_{k}$ operators can be formally solved using an approach similar to References \cite{ref20,ref15,ref16}:
\begin{align}
	r_{k}(t)&=r_{k}(t_{0})\text{e}^{-i\omega_{k}(t-t_{0})}-ig_{wg}\int_{t_{0}}^{t}p(u)\text{e}^{-i\omega_{k}(t-u)}du \nonumber \\
	l_{k}(t)&=l_{k}(t_{0})\text{e}^{-i\omega_{k}(t-t_{0})}-ig_{wg}\int_{t_{0}}^{t}p(u)\text{e}^{-i\omega_{k}(t-u)}du \nonumber
\end{align}
\normalsize
Substituting these formal solutions into the solutions of the cavity operator, one eventually finds
\begin{align}
\dot{p}&=-i\omega_{c}p(t)-ig_{wg}\left(\int r_{k} dk+\int l_{k} dk\right)- i\Omega S_{-} \nonumber \\
		&=-i\omega_{c}p(t)-ig_{wg}\left(\sqrt{2\pi}a_{input}(t)-\frac{2i\pi g_{wg}}{c}p(t)\right)- i\Omega S_{-} \nonumber \\
		&=-i\omega_{c}p(t)-\frac{2\pi g_{wg}^{2}}{c}p(t)-ig_{wg}\sqrt{2\pi}a_{input}(t)- i\Omega S_{-} \nonumber \\
		&=-i\omega_{c}p(t)-\frac{\kappa}{2}p(t)-i\sqrt{\frac{c\kappa}{2}}a_{input}(t)- i\Omega S_{-} \label{eqp}
\end{align}
\normalsize
whereby 
\begin{equation}
a_{input}(t)=\frac{1}{\sqrt{2\pi}}\int r_{k}(t_{0})\text{e}^{-i\omega_{k}(t-t_{0})} dk
\end{equation}
is the right propagating input field. There is no input field propagating to the left, hence $\int l_{k}(t_{0})\text{e}^{-i\omega_{k}(t-t_{0})}dk=0$. Similar to the results obtained in Reference \cite{ref20}, one can see from equation (\ref{eqp}) that the infinite waveguide degrees of freedom can be effectively integrated out. As shown in Reference \cite{ref20}, the dynamics of the overall system can then accurately be described by incorporating an additional Lindblad term to the master equation
\begin{equation}
 \sum_{\nu} \left(2 O_{\nu}\rho O_{\nu}^{\dagger} - O_{\nu}^{\dagger}O_{\nu}\rho - \rho O_{\nu}^{\dagger}O_{\nu}\right)
\end{equation}
with 
\begin{equation}
O_{\nu}= \sqrt{\frac{\kappa}{4}}p,\ \ \ \ \ \ \ \nu=\pm
\end{equation}
where $\nu$ distinguishes the right- and left-propagating fields. This additional Lindblad term
\begin{equation}
\frac{\kappa}{2}\left(2 p\rho p^{\dagger} - p^{\dagger}p\rho - \rho p^{\dagger}p\right)
\end{equation}
describes the cavity decay into the guided modes. In a frame rotating at $\omega_{L}$ the master equation then eventually becomes
\begin{multline}
\frac{d\rho}{dt}= -\frac{i}{\hbar}[\mathcal{H}_{rot},\rho] + \frac{\gamma_{p}}{2}\left(2 p\rho p^{\dagger} - p^{\dagger}p\rho - \rho p^{\dagger}p \right) \\ +\frac{\gamma_{e}}{2}\left(2 S_{-}\rho S_{+} - S_{+}S_{-}\rho - \rho S_{+}S_{-} \right)
\end{multline}
where
\begin{equation}
\mathcal{H}_{rot}=\mathcal{H}_{atom}+\mathcal{H}_{cavity}+\mathcal{H}_{coupling}+\mathcal{H}_{d}
\end{equation}
with
\begin{align}
\mathcal{H}_{atom} &= \hbar\left(\omega_{e}-\omega_{L}\right)S_{z} = \hbar\delta_{e}S_{z} \\
\mathcal{H}_{cavity} &= \hbar\left(\omega_{c}-\omega_{L}\right)p^{\dagger}p = \hbar\delta_{c}p^{\dagger}p \\
\mathcal{H}_{coupling} &=  \hbar \Omega \left(pS_{+}+p^{\dagger}S_{-}\right) \\
\mathcal{H}_{d} &= \hbar F p^{\dagger}+ \hbar \bar{F}p
\end{align}

\subsubsection{Numerical evaluation}
The first and second order correlation function can be evaluated using the steady-state ($t\rightarrow\infty$) density matrix $\rho^{ss}$ which is determined by $d\rho/dt=0$. We can determine the steady state solution by evaluating the master equation in a pre-defined basis
\begin{multline}
\{\ket{g,0}, \ket{g,1}, \ket{g,2}, \dots, \ket{g,N_{ex}},\ldots \\
\ldots\ket{e,0}, \ket{e,1}, \ket{e,2},\ldots, \ket{e,N_{ex}-1}\}
\end{multline}
which consists of the ground ($g$) and excited ($e$) states of the emitter and a certain number of cavity excitations such that the maximum number of excitations in the system is $N_{ex}$. If the emitter is in the ground state this means that the cavity can have $N_{ex}$ excitations, while the cavity can only have up to $N_{ex}-1$ excitations if the emitter is in the excited state. So for $N_{ex}$ excitations, the total number of basis states is $N_{b}=2N_{ex}+1$. These states will be numbered by $\{\ket{b},b=1\ldots N_{b}\}$, such that
\begin{multline}
\{\ket{1}=\ket{g,0}, \ldots, \ket{N_{ex}+1}=\ket{g,N_{ex}},\ldots \\ 
\ldots\ket{N_{ex}+2}=\ket{e,0}, \ldots, \ket{2N_{ex}+1}=\ket{e,N_{ex}-1}\}
\end{multline}
In order to solve the system, we recast the elements of the density matrix $\rho^{ss}_{\alpha\beta}, \alpha,\beta=1\ldots N_{b}$ into a a column vector of length $N_{b}^{2}$, i.e. $V^{\rho}_{j}, \{j=1\dots N_{b}^{2}\}$ and build up the Liouvillian superoperator $\mathcal{L}^{\rho}$, which is the $\left(N_{b}^{2}\times N_{b}^{2}\right)$ matrix representing all equations that determine $\rho^{ss}_{\alpha\beta}$, i.e.
\begin{equation}
\boxed{\sum_{j=1}^{N_{b}^{2}}\mathcal{L}^{\rho}_{ij}V_{j}^{\rho} = 0,\ \ \ \ \ \ \ \forall i=1\ldots N_{b}^{2}}
\end{equation}
The density matrix elements $\rho^{ss}$ are now determined by the eigenvector corresponding to the zero eigenvalue of $\mathcal{L}^{\rho}$. \cite{ref21} For all numerical evaluations we used a maximum number of excitations $N_{ex}=5$, so the total Hilbert space then consists of 11 basis states.

\subsubsection{First order correlation function}
The transmission ($\mathcal{T}$) and reflection ($\mathcal{R}$) coefficients are determined by the first order correlation function $g^{(1)}_{\mathcal{R},\mathcal{T}}$
\begin{equation}
g^{(1)}_{\mathcal{P}}=\sum_{m,n=1}^{N_{b}^{2}}\rho^{ss}_{nm}\langle m | a^{\dagger}_{\mathcal{P}}a_{\mathcal{P}}| n \rangle,\ \ \ \ \ \ \ \mathcal{P}=\{\mathcal{R},\mathcal{T}\}
\end{equation}
with
\begin{equation}
a_{\mathcal{P}}=a_{\mathcal{P}}^{i}+\psi p,\ \ \ \ \ \ \ a_{\mathcal{P}}^{i}=a_{IN}\delta_{\mathcal{P}\mathcal{T}},\ \ \ \psi=-i\sqrt{\frac{\kappa}{2c}}.
\end{equation}
such that
\begin{equation}
\boxed{
\mathcal{T}=\frac{g^{(1)}_{\mathcal{T}}}{a_{IN}^{2}},\ \ \ \ \ \ \ \mathcal{R}=\frac{g^{(1)}_{\mathcal{R}}}{a_{IN}^{2}}.
}
\end{equation}
The combination of the matrix elements $\langle m | a^{\dagger}_{\mathcal{P}}a_{\mathcal{P}}| n \rangle$ and $\rho^{ss}$ (calculated using the Liouvillian superoperator method as described before) fully determines the transmission and reflection coefficients (and of course should give the same outcome as the Heisenberg equation analysis).

\subsubsection{Second order correlation function}
The steady-state second order correlation function $g^{(2)}_{\mathcal{R},\mathcal{T}}(\tau=0)$ for the reflected  and transmitted field is determined by
\begin{equation}
g^{(2)}_{\mathcal{P}}(0)=\frac{\sum_{m,n=1}^{N_{b}^{2}}\rho^{ss}_{nm}\langle m | a^{\dagger}_{\mathcal{P}}a^{\dagger}_{\mathcal{P}}a_{\mathcal{P}}a_{\mathcal{P}} | n \rangle}{\left(g^{(1)}_{\mathcal{P}}\right)^{2}},\ \ \ \ \ \ \ \mathcal{P}=\{\mathcal{R},\mathcal{T}\}
\end{equation}

\subsection{Analytical approximation}
We derive an approximate analytical formula for $g^{(2)}_{\mathcal{R},\mathcal{T}}$ by limiting the Hilbert space to only 5 possible states, i.e. the maximum number of excitations in the system is $N_{ex}=2$ such that the possible states are $\{\ket{1}=\ket{g,0}, \ket{2}=\ket{g,1}, \ket{3}=\ket{g,2}, \ket{4}=\ket{e,0}, \ket{5}=\ket{e,1}\}$. For notational simplicity we note $\rho^{ss}=\rho$ furtheron. Using the earlier introduced notations, the numerator of $g^{(2)}_{\mathcal{P}}(0)$ ($\mathcal{P}=\{\mathcal{R},\mathcal{T}\}$) is then
\footnotesize
\begin{align}
&|a_{\mathcal{P}}^{i}|^{4}+4|a_{\mathcal{P}}^{i}|^{2}\Re\left(\overline{a_{\mathcal{P}}^{i}}\psi\left(\rho_{21}+\sqrt{2}\rho_{32}+\rho_{54}\right)\right) \nonumber \\
&\ \ \ +2\sqrt{2}\Re\left(\left(\overline{a_{\mathcal{P}}^{i}}\right)^{2}\psi^{2}\rho_{31}\right) +4|a_{\mathcal{P}}^{i}|^{2}|\psi|^{2}\left(\rho_{22}+2\rho_{33}+\rho_{55}\right) \nonumber \\
&\ \ \ \ \ \ \ +4\sqrt{2}|\psi|^{2}\Re\left(\overline{a_{\mathcal{P}}^{i}}\psi\rho_{32}\right)+2|\psi|^{4}\rho_{33}
\end{align}
\normalsize
Similarly, the denominator is given by the square of
\begin{multline}
g^{(1)}_{\mathcal{P}}=|a_{\mathcal{P}}^{i}|^{2}+2\Re\left(\overline{a_{\mathcal{P}}^{i}}\psi\left(\rho_{21}+\sqrt{2}\rho_{32}+\rho_{54}\right)\right) \\
+|\psi|^{2}\left(\rho_{22}+2\rho_{33}+\rho_{55}\right)
\end{multline}
In the weak probe limit we assume a sufficiently small coherent driving strength $F$ such that the expansion coefficients $c_{i}$ ($i=1\ldots 5$) in the wavefunction $\ket{\Psi}=\sum_{i=1}^{5}c_{i}\ket{i}$ are $c_{1}\approx 1$, $c_{2,4}=\mathcal{O}(F)$ and $c_{3,5}=\mathcal{O}(F^{2})$. This means that the cavity-emitter system is mostly in the groundstate $\rho_{11}\approx 1$ and that the one-excitation (2,4) and two-excitation (3,5) states are appropriately described by terms up to order one or two in $F$ respectively. As such we only keep terms up to order $F$ for $\rho_{12},\rho_{14}$, to order $F^{2}$ for $\rho_{13},\rho_{15},\rho_{22},\rho_{2,4},\rho_{44}$, to order $F^{3}$ for $\rho_{23},\rho_{25},\rho_{3,4},\rho_{45}$ and to order $F^{4}$ for $\rho_{33},\rho_{35},\rho_{55}$ in the master equation. The transmission and reflection coefficients are then determined by $\mathcal{P}=\frac{g^{(1)}_{\mathcal{P}}}{|a_{IN}|^{2}}$. For the transmission this means
\begin{align}
\mathcal{T}&\approx 1+\frac{2}{|a_{IN}|^{2}}\Re\left(\overline{a_{\mathcal{T}}^{i}}\psi\rho_{21}\right)+\frac{|\psi|^{2}}{|a_{IN}|^{2}}\rho_{22} \nonumber \\
&= 1+\frac{2|\psi|^{2}c^{2}}{F^{2}}\Re\left(\frac{-iF}{c}\rho_{21}\right)+\frac{|\psi|^{4}c^{2}}{F^{2}}\rho_{22} \nonumber \\
&= 1+\frac{\kappa}{F}\Im\left(\rho_{21}\right)+\frac{\kappa^{2}}{4F^{2}}\rho_{22} \nonumber
\end{align}
while the reflection coefficient is given by
\begin{equation}
\mathcal{R}=\frac{\kappa^{2}}{4F^{2}}\rho_{22}
\end{equation}
Introducing the density matrix elements as calculated by the simplified master equation yields the same reflection and transmission coefficient as the one obtained through the Heisenberg equation analysis, which serves as a sanity check of the applied approximations. In a similar way the $g^{(2)}_{\mathcal{T}}$ function is determined by
\begin{multline}
g^{(2)}_{\mathcal{T}}(0)\approx \frac{1}{\mathcal{T}^{2}}\left(1+\frac{2\kappa}{F}\Im\left(\rho_{21}\right)-\frac{\sqrt{2}\kappa^{2}}{2F^{2}}\Re\left(\rho_{31}\right)\right. \\
+\frac{\kappa^{2}}{F^{2}}\rho_{22}+\left.\frac{\sqrt{2}\kappa^{3}}{2F^{3}}\Im\left(\rho_{32}\right)+\frac{\kappa^{4}}{8F^{4}}\rho_{33}\right)
\end{multline}
and $g^{(2)}_{\mathcal{R}}$ is given by
\begin{equation}
g^{(2)}_{\mathcal{R}}(0)\approx \frac{\kappa^{4}}{8F^{4}\mathcal{R}^{2}}\rho_{33}=\frac{2\rho_{33}}{\rho_{22}^{2}}.
\end{equation}
The calculation of the density matrix elements $\rho_{mn}$ is straightforward but tedious and can be done with Mathematica. For non-zero detuning and $\delta_{e}=\delta_{c}=\delta$ the second order correlation function in transmission is given by 
\small
\begin{widetext}
\begin{multline}
g^{(2)}_{\mathcal{T}}(0)=\bigg\{\left[(4\Omega^2+\gamma_{e}\gamma_{p})^{2}+4\delta^{2}(\gamma_{e}^{2}+\gamma_{p}^{2}-8\Omega^{2})+16\delta^{4}\right]\times \\
\left[256\Omega^8+128\Omega^6\left(\gamma_{c}^{2}+2\gamma_{c}(\gamma_{e}+\kappa)-12\delta^{2}\right)+(\gamma_{c}^{2}+4\delta^{2})^{2}(\gamma_{e}^{2}+4\delta^{2})((\gamma_{e}+\gamma_{p})^{2}+16\delta^{2})+\right.\\ 
\left.8\Omega^{2}(\gamma_{c}^{2}+4\delta^{2})\left(\gamma_{e}\gamma_{c}(\gamma_{e}+\gamma_{p})(\gamma_{c}+2(\gamma_{e}+\kappa))+4\delta^{2}(\gamma_{c}(3\gamma_{e}-4\kappa)-(\gamma_{e}+\kappa)(3\gamma_{e}+2\kappa))-96\delta^{4}\right)\right.\\
\left.+16\Omega^{4}(\gamma_{c}^{2}(\gamma_{c}^{2}+2(\gamma_{e}+\kappa)(3\gamma_{e}+2\kappa)+\gamma_{c}(6\gamma_{e}+4\kappa))+8\delta^{2}(\gamma_{c}^{2}-5\gamma_{c}\gamma_{e}+(\gamma_{e}+\kappa)(\gamma_{e}+2\kappa))+208\delta^{4})\right]\bigg\} / \\
\bigg\{\left((4\Omega^2+\gamma_{e}\gamma_{c})^{2}+4\delta^{2}(\gamma_{e}^{2}+\gamma_{c}^{2}-8\Omega^{2})+16\delta^{4}\right)^{2}
(16\Omega^{4}+8\Omega^{2}(\gamma_{p}(\gamma_{e}+\gamma_{p})-8\delta^{2})+(\gamma_{p}^{2}+4\delta^{2})((\gamma_{e}+\gamma_{p})^{2}+16\delta^{2}))\bigg\}
\end{multline}
\end{widetext}
\normalsize
while in reflection we find
\small
\begin{widetext}
\begin{multline}
g^{(2)}_{\mathcal{R}}(0)=\frac{\left[\left(4\Omega^{2}+\gamma_{e}\gamma_{p}\right)^{2}+4\delta^{2}\left(\gamma_{e}^{2}+\gamma_{p}^{2}-8\Omega^{2}\right)+16\delta^{4}\right]\left[\left(\gamma_{e}\left(\gamma_{e}+\gamma_{p}\right)-4\Omega^{2}\right)^{2}+4\delta^{2}\left(16\Omega^{2}+5\gamma_{e}^{2}+2\gamma_{e}\gamma_{p}+\gamma_{p}^{2}\right)+64\delta^{4}\right]}{\left(\gamma_{e}^{2}+4\delta^{2}\right)^{2}\left[\left(4\Omega^{2}+\gamma_{p}\left(\gamma_{e}+\gamma_{p}\right)\right)^{2}+4\delta^{2}\left(\gamma_{e}^{2}+2\gamma_{e}\gamma_{p}+5\gamma_{p}^{2}-16\Omega^{2}\right)+64\delta^{4}\right]}.
\end{multline}
\end{widetext}

\end{document}